\documentclass{article}

\usepackage{arxiv}

\usepackage[utf8]{inputenc}
\usepackage[T1]{fontenc}
\usepackage{amsmath}
\usepackage{amssymb}
\usepackage{amsfonts}
\usepackage[numbers]{natbib}
\usepackage{doi}
\usepackage{siunitx}
\usepackage{hyperref}
\usepackage{url}
\usepackage{subcaption}
\usepackage{adjustbox}
\usepackage{array}
\usepackage{booktabs}
\usepackage{multirow}
\usepackage{verbatim}
\usepackage{graphicx}

\usepackage{rotating}
\usepackage{tabularx}
\usepackage{ragged2e}
\newcolumntype{Y}{>{\RaggedRight\arraybackslash}X}

\usepackage[linesnumbered,ruled,vlined]{algorithm2e}

\usepackage{tikz}
\usetikzlibrary{fit,calc}
\newcommand{\boxit}[2]{
    \tikz[remember picture,overlay] \node (A) {};\ignorespaces
    \tikz[remember picture,overlay]{\node[yshift=3pt,fill=#1,opacity=.25,fit={($(A)+(0,0.15\baselineskip)$)($(A)+(.8\linewidth,-{#2}\baselineskip - 0.25\baselineskip)$)}] {};}\ignorespaces
}

\PassOptionsToPackage{table,xcdraw}{xcolor}
\usepackage[normalem]{ulem}
\useunder{\uline}{\ul}{}

\title{KISS: Keeping it Simple and Slotted\\when Learning to Communicate over Wireless}

\renewcommand{\shorttitle}{KISS: Keeping it Simple and Slotted}

\hypersetup{
pdftitle={KISS: Keeping it Simple and Slotted when Learning to Communicate over Wireless},
pdfsubject={cs.NI, cs.LG},
pdfauthor={Kamil Szczech, Maksymilian Wojnar, Krzysztof Rusek, Katarzyna Kosek-Szott, Szymon Szott},
pdfkeywords={Distributed channel access, machine learning, MAC protocol},
colorlinks=false,
pdfborder={0 0 0},
}

\date{}
\renewcommand{\undertitle}{}
\renewcommand{\headeright}{}

\author{
  \href{https://orcid.org/0009-0002-4130-3086}{\includegraphics[scale=0.06]{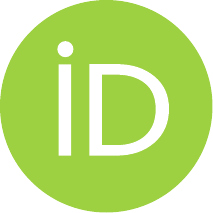}\hspace{1mm}Kamil Szczech},\quad
  \href{https://orcid.org/0000-0003-4510-5976}{\includegraphics[scale=0.06]{orcid.pdf}\hspace{1mm}Maksymilian Wojnar},\quad
  \href{https://orcid.org/0000-0003-4336-7841}{\includegraphics[scale=0.06]{orcid.pdf}\hspace{1mm}Krzysztof Rusek}
  \AND
  \href{https://orcid.org/0000-0001-7609-9420}{\includegraphics[scale=0.06]{orcid.pdf}\hspace{1mm}Katarzyna Kosek-Szott},\quad
  \href{https://orcid.org/0000-0001-5884-5581}{\includegraphics[scale=0.06]{orcid.pdf}\hspace{1mm}Szymon Szott} \\[1.5em]
  AGH University of Krakow, Poland \\
  \texttt{\href{mailto:kamil.szczech@agh.edu.pl}{kamil.szczech@agh.edu.pl}}
}

\begin{document}

\maketitle

\begin{abstract}
A long‑standing challenge in distributed wireless systems is ensuring efficient and fair random channel access. Existing solutions often address specific constraints related to timing, periodicity, or centralization, but they typically rely on fixed heuristics. Motivated by recent advances in machine learning (ML), we investigate whether ML agents can autonomously learn efficient and fair access strategies, and whether such learning can offer new insights into medium access control (MAC) design. Rather than proposing a deployable protocol, our aim is to examine whether decentralized learning can rediscover or approximate theoretically efficient random‑access mechanisms under minimal assumptions. To this end, we deploy an off‑policy Double Deep Q‑Network (DDQN) with Bayesian inference to train agents operating over a slotted channel. The resulting method is fully online (no pre‑training), fully distributed (independent multi‑agent learners), stochastic (non‑periodic), and requires no coordination or explicit communication. Extensive simulations show that the learned strategy adapts to varying network conditions and achieves near‑theoretical efficiency while maintaining fairness. Ablation studies further reveal that the learned behavior resembles slotted ALOHA with a dynamically adjusted transmission probability, leading us to refer to the method as KISS: Keeping It Simple and Slotted.
\end{abstract}


\section{Introduction}

Channel access protocols in distributed wireless systems originate from the original (pure) ALOHA protocol \cite{abramson1970aloha}.
Their evolution into more complex systems, through several improvements, such as slotted ALOHA and carrier sense multiple access (CSMA), has led to the enhanced distributed channel access (EDCA) function of IEEE 802.11, as well as to a multitude of protocols proposed in the literature \cite{sadeghi2017survey}.
Meanwhile, simpler protocols based on ALOHA remain relevant and used, e.g., in Internet of Things (IoT) \cite{balevi2018aloha,elkourdi2018enabling,seo2025aloha} and low-Earth orbit (LEO) satellite communications \cite{fidele2026analytical,tondo2025closeness}.

The operating limits of ALOHA-based protocols have been obtained analytically. For example, under full buffers, pure ALOHA and slotted ALOHA achieve throughput limits of $\frac{1}{2e} \approx 18.4\%$ and $\frac{1}{e} \approx 36.8\%$, respectively~\cite{abramson1977throughput}. For dynamic, time-varying settings, control‑theoretic approaches have shown how to optimally configure  the transmission probability~\cite{garcia2015adaptive}. With the rise of machine learning (ML) methods, \textbf{we ask whether ML agents can autonomously rediscover efficient behavior in fully distributed environments without explicit communication}. We apply learning agents primarily as an analytical tool to explore what channel‑access strategies can emerge under minimal assumptions, rather than as a deployable MAC protocol. This question remains pertinent because existing ML-based channel‑access methods typically rely on centralization, cyclic transmissions, or additional signaling (Section~\ref{sec:motivation}).

\begin{figure}[!t]
    \centering
    \includegraphics[width=0.35\linewidth]{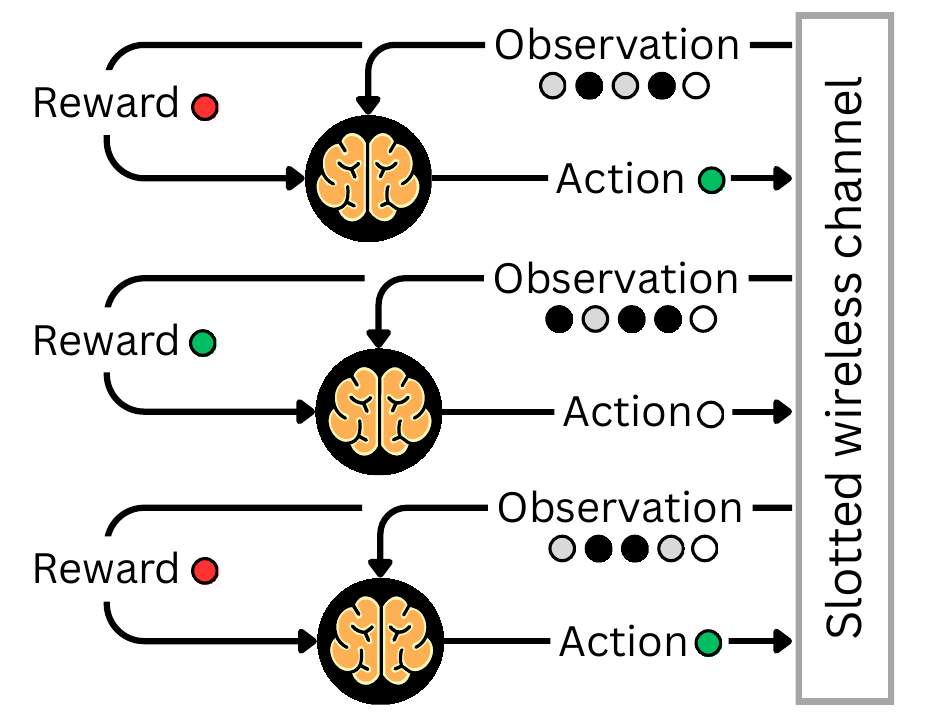}
    \caption{Simplified KISS operation diagram.}
    \label{fig:kiss}
\end{figure}

To address this challenge, we consider a distributed system composed of agents operating in synchronous time slots (Figure~\ref{fig:kiss}). Agents act independently, i.e., all learning and decision making is done locally. There is no central controller, no shared parameters, and no control messages. We formulate the problem as a partially observable stochastic game (POSG) and define the environment (i.e., slotted wireless channel) states, action space, and observations (Section~\ref{sec:system_model}).  We next design a reward function to promote efficient use of the channel, rewarding the agent for successful transmissions, penalizing for collisions, etc. We train agents using an off-policy Double Deep Q-Network (DDQN)~\cite{hasselt2016deep} with Bayesian inference. The training is done in a custom wireless channel access simulator (assuming a perfect physical layer), wherein we have implemented several ML and non-ML baselines (Section~\ref{sec:implementation}).

A thorough simulation study (Section~\ref{sec:performance}) shows that our agents approach the throughput limit of slotted ALOHA in full‑buffer scenarios. Baseline results indicate that higher throughput is only possible when relaxing the assumptions of distributed operation and no coordination. Despite these constraints, our agents maintain high fairness and adapt quickly in dynamic settings, converging to new operating points. In non‑saturated regimes, KISS is the only protocol that preserves both high throughput and fairness, outperforming all baselines under low and medium loads.
Ablation studies (Section~\ref{sec:ablation}) further reveal that KISS's performance stems primarily from the reward function, while modifications to the observation history or action‑space constraints offer little benefit and may even harm performance. Overall, the agents learn a simple yet efficient slotted‑ALOHA‑like strategy with dynamic transmission probabilities, which we term KISS: Keeping It Simple and Slotted.

In short, the contributions of this paper are as follows:

\begin{itemize}
    \item Defining limitations of existing ML‑based channel access schemes.
    \item Designing KISS, a fully distributed DDQN‑based access mechanism with zero coordination, zero shared parameters, and no prior structural assumptions.
    \item Re‑implementing state‑of‑the‑art baselines for reproducible comparisons.
    \item Demonstrating, through extensive experiments, that KISS achieves near‑optimal throughput, high fairness, and strong robustness across saturated, non‑saturated, static, and dynamic scenarios.
    \item Providing systematic ablation studies that identify reward shaping as the decisive factor enabling stable and fair decentralized operation.
    \item Releasing our simulator, baselines, and KISS implementation as open‑source.\footnote{\url{https://github.com/mldr-devs/ltc}}
\end{itemize}

\section{Motivation}
\label{sec:motivation}

Our literature review in the area of ML-based wireless channel access methods (Table~\ref{tab:survey}) reveals the following shortcomings which motivate our work.

\begin{table*}[!tp]
\centering
\footnotesize
\caption{Decentralized ML-based random-access methods in chronological order. Actions: transmit (TX), channel sense (CS). Requirements summarize additional assumptions beyond basic slot-level feedback.}
\label{tab:survey}
\begin{tabularx}{\linewidth}{
p{4em} p{3.2em} p{5em} p{4em} p{5em} p{4.5em} Y Y
}
\hline
\textbf{Paper} &
\textbf{Actions} &
\textbf{Observations} &
\textbf{Reward} &
\textbf{Mode} &
\textbf{Model} &
\textbf{Requirements} &
\textbf{Outcome vs. ALOHA}
\\ \hline

ALOHA-Q \cite{chu2012aloha} &
Slot choice &
Slot result &
$+1/-1$ &
Decentralized &
Q-learning &
Cyclic access; node count known &
Higher throughput and lower delay; converges to collision-free patterns.
\\

DLMA \cite{yu2019deep} &
TX, CS &
History of slot results &
$+1/0$ &
Decentralized/ centralized &
DDQN &
History buffer; centralized controller for fairness &
Maintains throughput and fairness with ALOHA nodes; not purely random access.
\\

UW-ALOHA-Q \cite{park2019reinforcement} &
Slot choice &
Slot result &
$+1/-1$ &
Decentralized &
Q-learning &
Signaling handshake; cyclic operation &
Improves underwater ALOHA; no terrestrial comparison.
\\

DR-ALOHA-Q \cite{tomovic2023dr} &
Slot choice &
Slot result &
$+1/0$ &
Decentralized &
Hysteretic Q-learning &
Cyclic operation &
Improves underwater performance; no terrestrial comparison.
\\

ALOHA-dQT \cite{zhang2022making} &
Cycle and slot choice &
Outcome/ACK history &
N/A &
Decentralized with info exchange &
Policy tree &
History and ACK summary exchange; coordinated cycles &
Higher utilization and fairness; relies on coordination beyond ALOHA.
\\

DMRL-MAC \cite{dutta2021towards} &
TX probability &
Local success/ collision rates &
Throughput-fairness metric &
Decentralized &
Hysteretic Q-learning &
Tracks per-node throughput; local monitoring &
Improved throughput under varying loads.
\\

DMLLI \cite{dutta2022distributed} &
TX probability &
Local success/ collision rates &
Throughput-fairness metric &
Decentralized with information exchange &
Hysteretic Q-learning &
Piggybacked throughput; neighborhood monitoring &
Performance gains when coexisting with ALOHA nodes.
\\

\textbf{KISS \break (this work)} &
TX, CS &
Channel outcome; local state &
See Sec.~\ref{sec:reward} &
Decentralized &
DDQN &
No signaling; no node count; slot-level feedback only &
Learns p-persistent ALOHA-like access; approaches $1/e$ throughput without coordination.
\\ \hline
\end{tabularx}
\end{table*}

\paragraph{Centralization.}
A substantial portion of the existing ML-based MAC literature depends on centralization either during operation or training. Many approaches adopt centralized training with distributed execution (CTDE), requiring a controller that aggregates experience or coordinates actions among nodes \cite{guo2022multi, chen2023scalable, xiao2023online, chen2025multitask, han2025foundation}. Even some decentralized designs become partially centralized when fairness is enforced: a central controller performs round‑robin scheduling and collects transmission outcomes from the agents \cite{yu2019deep}. Such dependency is incompatible with random‑access wireless settings where infrastructure may not be present or where nodes must operate autonomously. This motivates our focus on a fully distributed mechanism that does not rely on a coordinator, global knowledge, or centralized learning.

\paragraph{Cyclic transmissions.}
Several learning‑based ALOHA variants rely on structured, periodic, or frame‑based transmissions rather than the minimal assumptions of classic slotted random access. ALOHA‑Q and related underwater ALOHA adaptations operate over framed or cyclic structures \cite{chu2012aloha, chu2015application, park2019reinforcement, tomovic2023dr}, and thereby deviate from the simplicity of classical slotted ALOHA. Similarly, ALOHA-dQT intentionally learns coordinated periodic access patterns \cite{zhang2022making}, which, while improving channel utilization, introduces predetermined transmission cycles that reduce the spontaneity and independence expected from random-access systems. Our goal is instead to preserve the fundamental slotted nature of ALOHA without imposing periodic schedules or frame‑level structure.

\paragraph{Additional signaling.}
Many proposed solutions require explicit information exchange among nodes, increasing overhead and relying on signaling channels not available in minimal ALOHA-like systems. For example, ALOHA-dQT nodes exchange identifiers and summaries of past outcomes to coordinate expert policies \cite{zhang2022making}. Distributed approaches such as DMRL-MAC and DMLLI piggyback throughput information or observe all transmission outcomes \cite{dutta2021towards, dutta2022distributed}. Even when the underlying learning formulation is decentralized, the need for such side information contradicts the purely independent learning model where each node uses only its own local slot-level feedback. By contrast, our approach avoids any explicit signaling and assumes only the minimal observable: the channel outcome.

\paragraph{Network monitoring.}
Another recurring assumption is that nodes possess knowledge about the network state beyond what classic ALOHA requires. Some methods depend on knowing the number of competing nodes \textit{a priori} \cite{chu2012aloha, chu2015application}, while others require estimating or tracking the throughput of all agents \cite{dutta2021towards, dutta2022distributed}. Such monitoring effectively replaces random access with a form of distributed scheduling or coordinated adaptation. By contrast, we assume agents have no knowledge of network size, traffic demand, or other nodes' performance. This allows us to investigate whether a fully distributed deep reinforcement learning (RL) agent, operating solely on local slotted observations, can autonomously converge to classical p‑persistent ALOHA behavior.

\paragraph{Summary.} Although prior research demonstrates that reinforcement learning can improve or even surpass ALOHA in both homogeneous and heterogeneous settings, these gains typically rely on assumptions that depart from the simplicity of classic random access. Existing solutions often introduce centralization, cyclic transmission structures, explicit signaling, or require knowledge of global network parameters, all of which simplify learning but compromise decentralization. To our knowledge, no prior work shows that fully distributed deep reinforcement learning (DRL) agents can autonomously recover the classic p‑persistent ALOHA behavior ($\approx 1/e$ throughput) using only slot‑level feedback, i.e., without exchanging frames, sharing histories, or knowing the network size or topology. This gap motivates our central claim: \textit{to learn to communicate over wireless, it is enough to keep the structure simple and slotted}. Accordingly, we propose Keeping it Simple and Slotted (KISS), a fully distributed DDQN-based approach in which agents operate solely on the per-slot channel outcome. We demonstrate that this simplicity is sufficient for convergence toward theoretically optimal behavior.

\section{System Model}
\label{sec:system_model}

We consider a distributed system composed of $N$ independent agents operating in synchronous time slots $t=0,1,\dots, \allowbreak T-1$ (Algorithm~\ref{alg:kiss}). The shaded box in Algorithm~\ref{alg:kiss} highlights the sequence of local operations executed independently by each agent (these steps run in parallel on all active agents). Agents act independently, i.e., learning and decision making processes run locally on each node: there is no central controller, no shared parameters, and no control messages.

\begin{center}
\begin{minipage}{0.55\linewidth}
\begin{algorithm}[H]
\small
\caption{KISS: Distributed Learning Loop.}
\label{alg:kiss}
\SetKwInput{Init}{Initialization}
\Init{Agents $i \in \{1..N\}$ with weights $\theta_i$, history $o_0^i = \mathbf{0}$}
\For{each time slot $t = 0, 1, \dots, T-1$\label{alg:main_loop}}{
    \For{each active agent $i \in \{1, \dots, N\}$ \textbf{in parallel}}{
        \boxit{blue!20}{6}
        Observe $x_t^i$ (\ref{eq:instant_obs}) and update history $o_t^i$ (\ref{eq:obs})\label{alg:agent_start}\;
        Calculate reward $r_{t-1}^i$ based on (\ref{eq:reward_tx}) and (\ref{eq:reward_cs})\label{alg:reward}\;
        Store transition $(s_{t-1}^i, a_{t-1}^i, r_{t-1}^i, s_t^i)$ in replay buffer $\mathcal{D}_i$\label{alg:store}\;
        Update weights $\theta_i$ by minimizing total loss $\mathcal{L}$ (\ref{eq:total_loss})\label{alg:update}\;
        Sample Bayesian parameter sample $\tilde{\theta}_i$ (\ref{eq:sampling})\label{alg:sample}\;
        Sample next action $a_t^i$ using $\epsilon$-greedy policy\label{alg:agent_end}\;
    }
    Collect and execute all actions simultaneously\label{alg:execute}\;
    Determine channel state (Section~\ref{sec:simulator})\label{alg:sim}\;
    Provide observations to agents\label{alg:reward_line}\;
    Update buffer states and generate new traffic (Section~\ref{sec:simulator})\label{alg:traffic}\;
}
\end{algorithm}
\end{minipage}
\end{center}

\subsection{Problem Formulation}

The problem can be formulated as a partially observable stochastic game (POSG) $(\mathcal{I},\mathcal{S},\{\mathcal{A}_i\},P,\{R_i\},\{\Omega_i\},\{O_i\},\gamma)$, where $\mathcal{I}=\{1,\dots,N\}$ is the set of agents, $\mathcal{S}$ is a set of environment states, $\mathcal{A}_i$ is the action set of agent $i$, $P(s'|s,\mathbf{a})$ is a transition probability conditioned on the joint action $\mathbf{a}=(a^1,\dots,a^N)$, $R_i:\mathcal{S}\times\prod_i\mathcal{A}_i\to\mathbb{R}$ is the reward function of agent $i$, $\Omega_i$ is the observation set of agent $i$, $O_i(o^i|s',\mathbf{a})$ is the conditional observation probability for agent $i$, and $\gamma\in[0,1]$ is the discount factor. The global state space $\mathcal{S}$ is hidden from the agents, while $P$ and $O_i$ are governed by the simulator as detailed in Section~\ref{sec:implementation}.

Each agent $i\in\{1,\dots,N\}$ at each slot selects one action from the set
\begin{equation}
    \mathcal{A}=\{\mathrm{TX},\ \mathrm{CS}\},
    \label{eq:actions}
\end{equation}
corresponding to transmit (TX) or channel sense (CS). Note that there are no restrictions on action selection, therefore an agent is allowed to take any action at any state, even if that behavior is unfavorable (e.g., transmission with empty buffer). The local instantaneous (per-slot) observation at time $t$ is a 5-tuple
\begin{equation}
    x_t^i = \big(b_t^i,\; c_t^i,\; r_{c,t}^i,\; n_{\mathrm{idle},t}^i,\; a_t^i\big),
    \label{eq:instant_obs}
\end{equation}
where $b_t^i\in\{0,1\}$ is the binary buffer occupancy state (empty, full), $c_t^i\in\{-1,0,1\}$ is the channel state observed in the previous slot (undetermined/collision, idle, successful transmission), $r_{c,t}^i\in\mathbb{Z}_{\ge0}$ is the retransmission counter, $n_{\mathrm{idle},t}^i\in\mathbb{Z}_{\ge0}$ is the number of consecutive idle slots, and $a_t^i\in\{\mathrm{TX},\ \mathrm{CS}\}$ is the action taken at time $t$.
When the agent chose $\mathrm{CS}$ at time $t-1$, the channel state is observed as $c_t^i = 0$ (idle), $c_t^i = 1$ (successful transmission), or $c_t^i = -1$ (collision). When the agent chose $\mathrm{TX}$, it cannot sense the channel, so $c_t^i$ is set to $-1$ (undetermined). The observation provided to the agent is the history of the last $L$ instantaneous observations:
\begin{equation}
    o_t^i = \big[x_{t-L+1}^i,\; x_{t-L+2}^i,\; \dots,\; x_t^i\big],
    \label{eq:obs}
\end{equation}
therefore $\Omega \subseteq \mathbb{Z}^{L\times 5}$.

\subsection{Reward Function}
\label{sec:reward}

The reward function is designed to promote efficient use of the channel. It encourages successful packet delivery with minimal retransmissions, discourages wasteful transmissions with empty buffers or collisions, and penalizes excessive waiting to send packets.

\paragraph{Transmission:}
When the agent selects TX, let $\tau\in\{-1,1\}$ denote the transmission outcome\footnote{Following other state-of-the-art papers, we  assume that acknowledgments are sent back immediately and their transmission is part of the slot duration.} ($\tau=1$ for success, $\tau=-1$ for collision) and $b$ the buffer state. The reward is
\begin{equation}
r_{\mathrm{TX}} = \begin{cases}
R_{\mathrm{tx}},   & \tau=1,\; b>0, \\
P_{\mathrm{empty}},                 & \tau=1,\; b=0, \\
P_{\mathrm{coll}},                  & \tau=-1,\; r_c < M, \\
P_{\mathrm{max}},                  & \tau=-1,\; r_c \ge M,
\end{cases}
\label{eq:reward_tx}
\end{equation}
where $r_c$ is the retransmission counter and $M$ is the maximum number of retransmissions (Table~\ref{tab:constants}). In all cases, $n_{\mathrm{idle}}$  is set to zero. The $r_c$ counter is reset to zero on successful or empty buffer transmission. On collision, $r_c$ is incremented; if it reaches $M$, the frame is dropped and $r_c$ is reset.

\paragraph{Channel sense:}
When the agent selects CS, the reward depends on the buffer state and idle duration:
\begin{equation}
r_{\mathrm{CS}} = \begin{cases}
R_{\mathrm{idle}},   & b=0, \\
0,                    & b>0,\; n_{\mathrm{idle}} < s, \\
P_{\mathrm{idle}} \cdot \min\!\big(1,(n_{\mathrm{idle}} - s + 1)/D\big), & b>0,\; n_{\mathrm{idle}} \ge s,
\end{cases}
\label{eq:reward_cs}
\end{equation}
where $s = S_{\mathrm{idle}} + \lfloor\xi\rceil$, $\xi \sim \mathcal{N}(0, \sigma_{\mathrm{idle}}^2)$, $\lfloor\cdot\rceil$ denotes rounding to the nearest integer, and $D$ is the penalized idle scaling factor. When the buffer is empty, both counters are reset to zero. Otherwise, the $r_c$ counter remains unchanged and $n_{\mathrm{idle}}$ is incremented.
After the grace-period threshold is exceeded, the penalty for waiting with a non-empty buffer increases and then saturates at a fixed maximum value.

\begin{table}[t]
\caption{Reward function constants.}
\label{tab:constants}
\centering
\small
\begin{tabular}{lcc}
\toprule
\textbf{Constant} & \textbf{Symbol} & \textbf{Value} \\
\midrule
Max.\ retransmissions & $M$ & 8 \\
Safe idle period & $S_{\mathrm{idle}}$ & 25 \\
Safe idle std.\ dev. & $\sigma_{\mathrm{idle}}$ & 3 \\
Penalized idle period & $D$ & 25 \\
\specialrule{0.4pt}{2pt}{2pt}
Successful transmission reward & $R_{\mathrm{tx}}$ & 1.0 \\
Empty buffer idle reward & $R_{\mathrm{idle}}$ & 0.5 \\
Idle penalty & $P_{\mathrm{idle}}$ & $-$1.0 \\
Empty buffer TX penalty & $P_{\mathrm{empty}}$ & $-$0.5 \\
Collision penalty & $P_{\mathrm{coll}}$ & $-$1.0 \\
Max.\ retransmission penalty & $P_{\mathrm{max}}$ & $-$1.0 \\
\bottomrule
\end{tabular}
\end{table}

\subsection{Learning scheme and neural network model}

KISS agents follow a DDQN algorithm with a Q-network parameterized by $\theta$ and a target network by $\theta^{-}$. The target network is updated using a soft update rule. The DDQN bootstrap target is
\begin{equation}
    y = r + (1 - \delta) \gamma\, Q_{\theta^-}\!\big(s', \operatorname*{argmax}_{a'} Q_{\theta}(s',a')\big),
    \label{eq:bootstrap}
\end{equation}
and the standard temporal-difference (TD) loss is
\begin{equation}
    \mathcal{L}_{\mathrm{TD}}(\theta)=\mathbb{E}_{(s,a,r,s',\delta)\sim\mathcal{D}}\big[\,\big(Q_{\theta}(s,a)-y\big)^2\,\big],
    \label{eq:td_loss}
\end{equation}
where $\mathcal{D}$ denotes the replay buffer and $\delta$ is a terminal flag. Moreover, KISS agents use $\varepsilon$-greedy exploration with an exponential decay factor.

The Q-network architecture is a one-layer transformer classifier~\cite{vaswani2017attention} with a final fully connected layer producing the action logits. The choice of the transformer model is motivated by the fact that, to achieve distributed channel access with decentralized operation, an ML solution is needed that can handle variable‑length sequences (i.e., different history sizes)\footnote{While the main experiments use a fixed history length $L=10$, the transformer architecture enables evaluating different values of $L$ without modifying the network structure. We exploit this property in the ablation study (Section~\ref{sec:ablation}), where we compare $L=1$ and $L=10$ using the same model.}.
The Q-network is implemented as a Bayesian neural network~\cite{jospin2022handson} and each parameter is represented by a Gaussian posterior with diagonal covariance (mean field approximation).
For each parameter, we sample a noisy realization for the forward pass:
\begin{equation}
    \tilde{\theta}_i = \theta_i + \sigma_i\xi,\qquad \xi\sim\mathcal{N}(0,1),
    \label{eq:sampling}
\end{equation}
where $\sigma_i$ is a learnable standard deviation and $\xi$ is sampled from a standard normal distribution.
We evaluate a Monte Carlo estimate of the Kullback--Leibler (KL) divergence term
\begin{equation}
    \mathrm{KL}\approx\sum_{i}\Big(\log\varphi\big(\tilde{\theta}_i;\theta_i,\sigma_i\big) - \log\varphi\big(\tilde{\theta}_i; 0,\sigma_p\big)\Big),
    \label{eq:kl}
\end{equation}
where $\varphi$ is the normal distribution density, $\sigma_p$ is a prior standard deviation, and the summation is over all dimensions of all parameters. The total loss minimized during training is therefore
\begin{equation}
    \mathcal{L}(\theta) = \mathcal{L}_{\mathrm{TD}}(\theta) \;+\; \lambda_{\mathrm{KL}}\,\mathrm{KL}(\theta),
    \label{eq:total_loss}
\end{equation}
where $\lambda_{\mathrm{KL}}$ is a regularization coefficient. The full set of training hyperparameters is listed in Table~\ref{tab:hyperparams}.

This stochastic network allows injecting data‑driven noise into the learned protocol (the learning algorithm can cancel this noise by setting an arbitrarily small $\sigma_i$). This noise captures the agent's uncertainty due to unseen configurations, and in the multi‑agent setting it gives agents the option to use mixed strategies. This option is beneficial, as the Nash equilibrium often lies in a mixed strategy.

\begin{figure}[!t]
    \centering
    \begin{minipage}[t]{0.48\linewidth}
        \includegraphics[width=\linewidth]{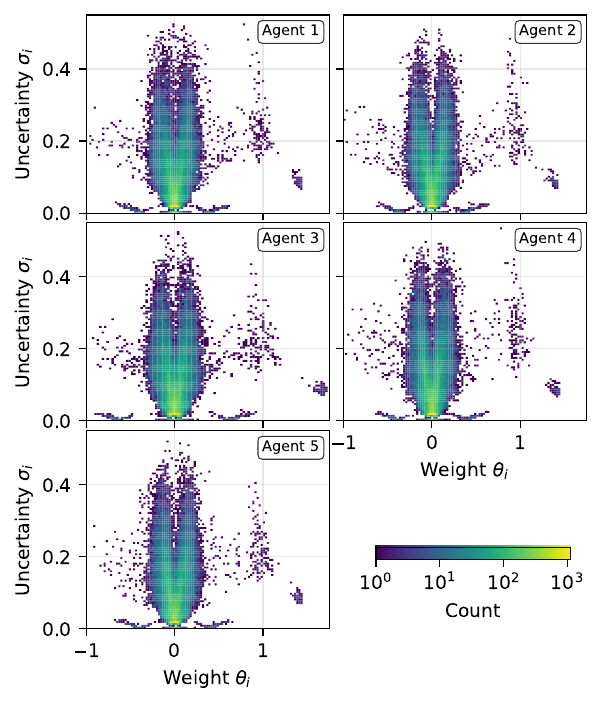}
        \caption{Joint distribution of weight means $\theta_i$ and the uncertainty measured by the standard deviations $\sigma_i$ for each of the five agents from a single training run ($N=5$, saturated traffic).}
        \label{fig:svi-hist2d}
    \end{minipage}
    \hfill
    \begin{minipage}[t]{0.48\linewidth}
        \includegraphics[width=\linewidth]{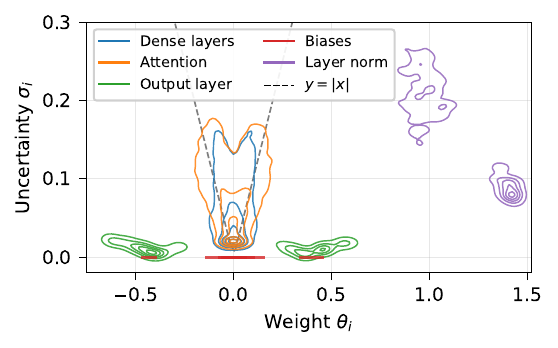}
        \caption{Joint distribution of weight means $\theta_i$ and uncertainty measured by the standard deviations $\sigma_i$ for Agent~1, shown as a contour plot and colored by layer category. Biases are fixed at zero by design.}
        \label{fig:svi-scatter}
    \end{minipage}
\end{figure}

To illustrate the learned posteriors, we examine a single representative run with $N=5$ agents under saturated traffic. Figure~\ref{fig:svi-hist2d} shows the joint distribution of weight means $\theta_i$ and posterior standard deviations $\sigma_i$ for each agent. A characteristic V-shape emerges: weights near zero are learned with low posterior variance, while larger-magnitude weights carry proportionally higher $\sigma_i$. All five agents develop nearly identical distributional shapes despite fully independent training, confirming that stochastic variational inference (SVI) yields symmetric stochastic policies without explicit coordination.

Figure~\ref{fig:svi-scatter} decomposes the posterior of Agent~1 by layer category. Dense and attention layer weights constitute the majority of parameters and form the V-shaped envelope. The output layer weight converges to near-zero $\sigma_i$, meaning the final mapping from features to action logits is essentially deterministic. Biases are fixed at zero by design and are not subject to SVI. Layer normalization parameters settle at larger positive $\theta_i$ with low variance. This reveals that the stochasticity originates in the internal feature representations rather than the output head -- the agents randomize \emph{how} they perceive the channel state, not the decision rule applied to a given representation.

\begin{table}[t]
\caption{DDQN training hyperparameters.}
\label{tab:hyperparams}
\centering
\small
\begin{tabular}{lc}
\toprule
\textbf{Parameter} & \textbf{Value} \\
\midrule
\multicolumn{2}{c}{\textit{Q-network architecture}} \\
Transformer layers & 1 \\
Hidden dimension & 64 \\
Attention heads & 4 \\
Prior scale ($\sigma_p$) & 2.0 \\
$\lambda_{\mathrm{KL}}$ & 0.1 \\
\specialrule{0.4pt}{2pt}{2pt}
\multicolumn{2}{c}{\textit{Optimizer (Adam + cosine decay)}} \\
Learning rate (init $\to$ final) & $10^{-4} \to 10^{-6}$ \\
Decay steps & 60\,000 \\
Gradient clipping & 1.0 \\
$\beta_1,\; \beta_2$ & 0.95,\; 0.95 \\
\specialrule{0.4pt}{2pt}{2pt}
\multicolumn{2}{c}{\textit{Reinforcement learning}} \\
Discount factor ($\gamma$) & 0.95 \\
Replay buffer size & 30\,000 \\
Batch size & 128 \\
Training steps per env.\ step & 5 \\
Target network $\tau$ & 0.05 \\
$\varepsilon$ (init $\to$ min) & $1.0 \to 0.0$ \\
$\varepsilon$ decay & 0.999 \\
\bottomrule
\end{tabular}
\end{table}

\section{Implementation}
\label{sec:implementation}

We now describe the environment in which the agents learn, our network and traffic assumptions, and define the metrics used to evaluate agent performance.

\subsection{Simulator}
\label{sec:simulator}
The network simulator is developed in the JAX framework, which accelerates simulation execution. The simulator is based on a slotted (Monte Carlo) model, which is a key assumption of the system. The slot duration is not tied to any specific unit of time. All slots are considered equal, and within each slot, one of three outcomes may occur: a successful packet transmission, a collision, or the channel may be idle.

Our MAC simulator design relies on several simplifications. First, the channel model does not account for signal attenuation with distance and there are no channel errors. Transmission errors in the radio channel occur only when multiple stations transmit simultaneously. Second, all stations are located within their transmission ranges, eliminating the problem of hidden stations.

The simulator follows Algorithm~\ref{alg:kiss}. It iterates over consecutive slots and, in each slot, first determines the channel status based on the actions selected by the agents. If more than one agent chooses to transmit, the simulator registers a collision and all colliding transmissions fail. Next, the simulator updates each agent’s buffer: if an agent transmitted and no collision occurred, its buffer is cleared. The buffer is then incremented according to the output of the traffic generator.

We model the traffic with a double‑stochastic process. For each transmission opportunity (TXOP), we first sample the rate from a lognormal distribution and then draw the number of frames from an independent Poisson distribution with that sampled rate. This approach provides flexibility, allowing the variance and mean rate to be controlled independently.

Afterwards, the simulator updates the observation histories with new data and removes the oldest entries. Finally, it computes the rewards for each agent independently.

\subsection{Baselines}
We consider the following slotted channel access protocols as our baselines for comparison, including  both traditional and learning-based methods:
\begin{itemize}
    \item Exponential backoff ALOHA (EB-ALOHA) -- we select an ALOHA variant as a non-ML baseline. The simplest variant, which uses a fixed transmission probability, is unsuitable in our evaluation scenarios where the number of nodes is unknown \textit{a priori}. Therefore, we use the more adaptable EB-ALOHA. Following the settings of \cite{yu2019deep}, we use an initial window size of 4 and a maximum backoff stage of 2, corresponding to a contention window range of $[4, 16]$. This baseline can be viewed as a rough approximation of Wi‑Fi's distributed coordination function (DCF), whose contention window range is larger but whose operation is unslotted.
    \item ALOHA-Q \cite{chu2012aloha} -- a pioneering RL-based enhancement of ALOHA, which assumes cyclic transmission patterns. We consider two configurations: (i) \textit{ideal}, in which the transmission cycle is perfectly (though unrealistically) aligned with the actual number of stations, and (ii) \textit{fixed}, where the transmission cycle is set to 10, unless otherwise specified.
    \item Deep reinforcement Learning Multiple Access (DLMA) \cite{yu2019deep} -- a widely cited DRL approach designed to maximize throughput or fairness, with a strong focus on coexistence with other access methods. DLMA operates in a fully decentralized manner only for throughput-oriented optimization. For fairness optimization, it requires a centralized controller implementing, e.g., simple round-robin scheduling. Hence, in our evaluation, we only use the throughput-oriented variant.
\end{itemize}
In summary, EB‑ALOHA serves as a non-ML reference point, while ALOHA‑Q and DLMA represent state‑of‑the‑art RL‑driven approaches. The remaining works listed in Table~\ref{tab:survey} rely on assumptions that differ substantially from ours, rendering them incompatible for direct comparison.

\subsection{Metrics}
We evaluate the performance of KISS and the aforementioned baselines using two complementary metrics: aggregate throughput and fairness.

First, we consider normalized aggregate network throughput: the ratio of slots with successful transmissions to the total number of slots.
This metric represents channel utilization.
Ideally, a utilization of $1.0$ is achievable only in fully centralized systems or systems with cyclic transmissions (such as the ideal ALOHA-Q), where the number of cycles exactly matches the number of active stations.
Meanwhile, the theoretical maximum utilization for a distributed slotted ALOHA system is $\frac{1}{e} \approx 36.8$\% \cite{abramson1977throughput}.
Consequently, reaching this bound represents the best achievable performance for our distributed KISS agents.

However, relying solely on throughput can be misleading because in a distributed system all participants should receive an equal share of transmission opportunities.
We want to avoid cases where a single dominant node transmits continuously, at the expense of others.
Therefore, we also study fairness in channel utilization by measuring Jain's fairness index:
\begin{equation}
  J(x_1, x_2, \ldots, x_n) = \frac{\left( \sum_{i=1}^{n} x_i \right)^2}{n \cdot \sum_{i=1}^{n} x_i^2}
\end{equation}
where $x_i$ represents the number of successful transmissions by station $i$, and $n$ is the number of stations.
The goal of distributed access methods is to achieve values close to $1.0$ (perfect fairness).

\section{Performance Analysis}
\label{sec:performance}
We evaluate KISS against the defined baselines in saturated networks, considering both instantaneous behavior for a 15‑station system and steady‑state scalability across different network sizes. We then analyze performance under non‑saturated low‑ and medium‑load traffic profiles, followed by a dynamic scenario where stations join or leave the network. Together, these experiments assess throughput, fairness, convergence time, and adaptability.
All experiments, including the ablation studies in Section~\ref{sec:ablation}, were repeated five times with different random seeds. All plots show the mean performance, with shaded regions indicating the 95\% confidence interval.

\subsection{Instantaneous Case}

\begin{figure}[!t]
    \centering
    \begin{minipage}[t]{0.48\linewidth}
        \includegraphics[width=\linewidth]{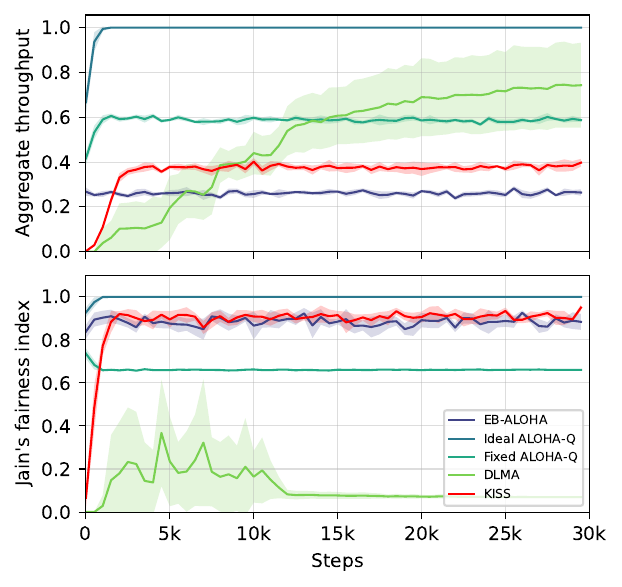}
        \caption{Performance for the instantaneous case.}
        \label{fig:Instantaneous_thr_fair}
    \end{minipage}
    \hfill
    \begin{minipage}[t]{0.48\linewidth}
        \includegraphics[width=\linewidth]{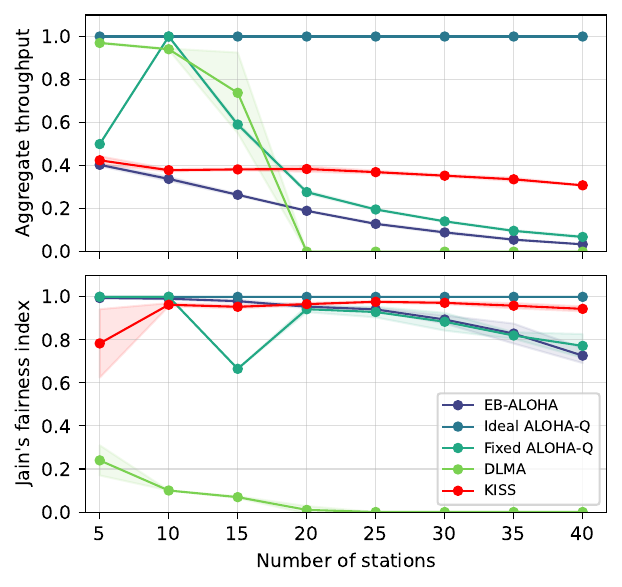}
        \caption{Performance for the steady-state case.}
        \label{fig:steadystate}
    \end{minipage}
\end{figure}

In our first study, the instantaneous case, we analyze the performance of each protocol for a network of 15 stations by measuring how each station's throughput evolves over time and how long it takes the protocol to converge to a stable solution. Figure~\ref{fig:Instantaneous_thr_fair} presents the performance of the protocols over time. EB‑ALOHA immediately reaches stable performance in both throughput and fairness. In contrast, RL‑based protocols require an exploration phase before stabilizing. The two ALOHA‑Q variants converge the fastest, while DLMA requires the longest exploration time; even after 30,000 steps, it still does not reach a stable state. KISS converges significantly faster than DLMA but more slowly than the ALOHA‑Q protocols. This is expected: ALOHA‑Q operates under simpler assumptions, making it easier for the agent to maximize its reward. Finding a suitable slot within a known interval is far simpler than discovering a new, optimal transmission strategy from scratch.

EB‑ALOHA achieves the lowest aggregate throughput, remaining below~0.3. ALOHA‑Q maintains stable throughput after its exploration phase. For the ideal (but unrealistic) ALOHA‑Q, throughput reaches~1. However, when the number of slots is limited to~10 (fixed ALOHA‑Q), throughput drops significantly (to~0.6). This reduction stems from having too few slots for a network of 15 stations (only 66.7\% of stations can transmit simultaneously) and larger networks experience even greater degradation. A similar trend appears in the fairness results: the fairness drop for fixed ALOHA‑Q indicates unequal channel access.

DLMA's  throughput keeps rising, indicating that the network remains in the exploration phase. The large standard deviation reflects learning instability, and fairness declines over time, suggesting emerging selfish behavior in which one agent continues transmitting while others stay idle to avoid collision penalties.

Meanwhile, our solution (KISS) achieves higher throughput than EB‑ALOHA, maintaining a value of approximately~0.4. KISS also preserves high fairness, only slightly lower than that of ALOHA‑Q. Thus, despite operating in a fully distributed manner, KISS agents maintain fair channel access and effectively prevent selfish behavior. These results confirm the superiority of KISS over the other channel‑access schemes.

\subsection{Steady-state Case}
For the steady‑state case, we evaluate the scalability of each protocol by testing its performance in networks of different sizes (Figure~\ref{fig:steadystate}). The ideal ALOHA‑Q maintains the maximum possible throughput and fairness regardless of the number of active stations. In contrast, fixed ALOHA‑Q achieves optimal throughput and fairness only when the number of available time slots matches the number of stations (i.e., 10). In all other cases, the larger the mismatch between the number of stations and slots, the lower the resulting throughput.

We observe a significant drop in fairness when the system consists of 15 stations. At this stage, 10 stations already identify and stabilize in their optimal time slots, while the remaining 5 are still looking for suitable ones. The problem is that these 10 stations have entered the exploitation phase, whereas the other 5 are still in the exploration phase. This imbalance leads to unfair resource allocation. When the number of stations is larger, this issue does not occur because there are too few time slots available. As a result, all stations are forced to continuously explore and none of them can fully settle into pure exploitation.

EB-ALOHA, similarly to fixed ALOHA-Q, faces scalability issues. As the number of stations increases, both the aggregate throughput and the fairness index decrease.

DLMA, in the range of 2 to 15 stations,  achieves high throughput but suffers from low fairness. For larger networks, its throughput drops to zero. Therefore, both selfish behavior (small networks) and protocol collapse (20+ stations) can be observed.

\begin{figure}[!t]
    \centering
    \includegraphics[width=0.5\linewidth]{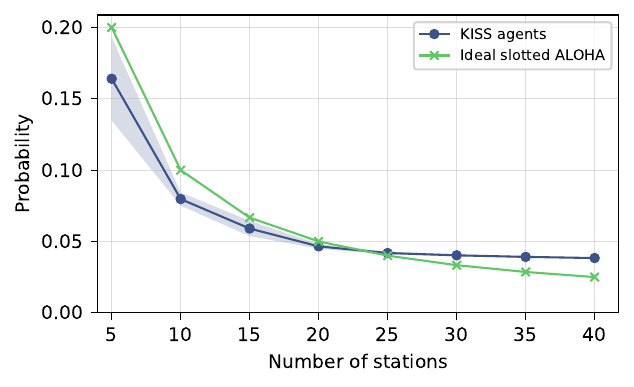}
    \caption{Comparison of the transmission probability of KISS agents versus ideal slotted ALOHA ($1/N$).}
    \label{fig:tx_prob_curve}
\end{figure}

KISS maintains an aggregate throughput of $\approx0.4$ and high fairness across the entire station range, which approximates the slotted ALOHA throughput limit of $1/e$~\cite{garcia2015adaptive}. These results indicate that KISS is scalable in networks with an unknown number of transmitters and considerably outperforms other non-ideal solutions. Furthermore, Figure~\ref{fig:tx_prob_curve} shows the transmission probability chosen by the KISS agent as a function of the number of stations, compared to the theoretical optimum of $1/N$. The agents' probability closely follows the optimal curve across the entire range, demonstrating that KISS adaptively approximates the ideal transmission probability without any explicit knowledge of the number of active stations.

\subsection{Non-saturated Traffic Profiles}

\begin{figure}[!t]
    \centering
    \begin{minipage}[t]{0.48\linewidth}
        \includegraphics[width=\linewidth]{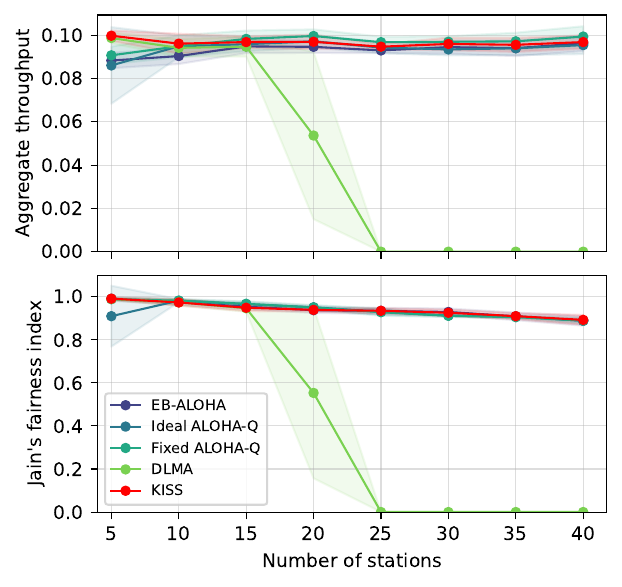}
        \caption{Steady-state case under low traffic.}
        \label{fig:steadystate-low}
    \end{minipage}
    \hfill
    \begin{minipage}[t]{0.48\linewidth}
        \includegraphics[width=\linewidth]{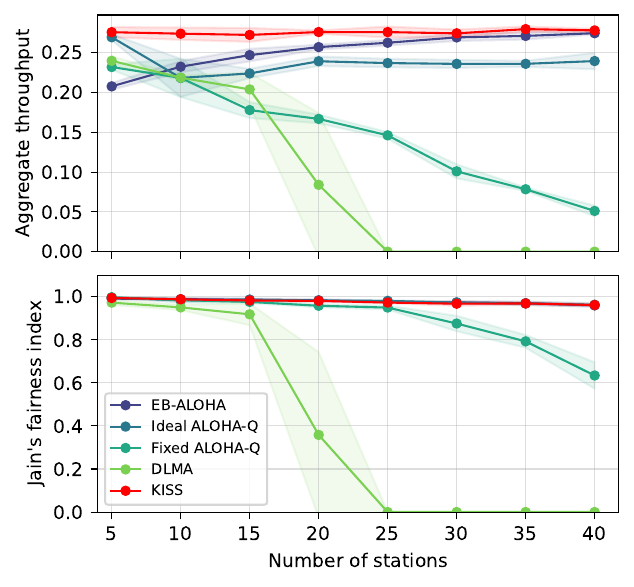}
        \caption{Steady-state case under mid traffic.}
        \label{fig:steadystate-mid}
    \end{minipage}
\end{figure}

We now examine how the protocols behave under non‑saturated network conditions --specifically, low and medium traffic. The station traffic generators produce enough data for the aggregate load to reach at most $0.1$ in the low‑traffic case and $0.3$ in the medium‑traffic case. As shown in Figure~\ref{fig:steadystate-low}, all protocols except DLMA perform similarly in the low‑traffic scenario, regardless of network size. The aggregate throughput remains close to $0.1$, matching the offered load, and the fairness index stays near~1. DLMA achieves comparable performance for network sizes between 1 and 15 stations but degrades rapidly as the network grows larger.

\begin{figure}[!t]
    \centering
    \includegraphics[width=0.5\linewidth]{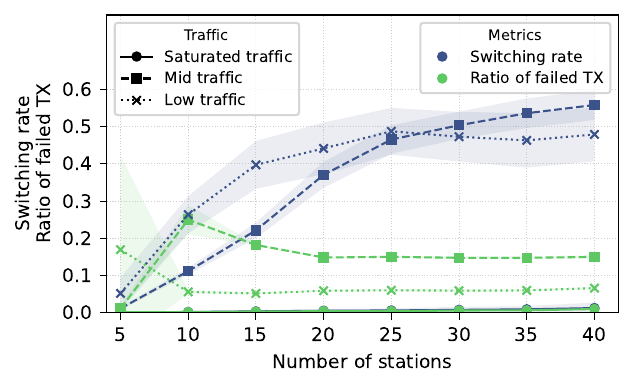}
    \caption{ALOHA-Q slot switching rate and ratio of failed transmissions vs no. of stations.}
    \label{fig:switch-fail-tx}
\end{figure}

Under medium traffic conditions (Figure~\ref{fig:steadystate-mid}), only KISS maintains a stable aggregate throughput (of $\approx0.3$). Both DLMA and fixed ALOHA-Q lose efficiency as the number of stations increases. Furthermore, EB-ALOHA achieves a lower throughput than KISS in the range of 1-20 stations; beyond this range, its throughput is comparable to that of KISS.
Interestingly, the ideal ALOHA-Q maintains a throughput of $0.25$. This is because the reward in ALOHA-Q is updated only when a data frame is sent. With low network load, reward updates are rare, which makes the learning process inefficient. As a result, agents cannot find permanent slots to carry out stable transmissions. This instability limits the system from achieving the maximum aggregate throughput possible in a reasonable time.
To better explain the observed behavior, Figure~\ref{fig:switch-fail-tx} illustrates the ALOHA-Q slot switching rate and the ratio of failed transmissions. The figure presents the mean values together with confidence intervals obtained from five independent agents in each experiment. The slot switching rate measures how frequently ALOHA-Q agents change their selected transmission slot; higher values indicate more frequent slot switching.
In addition, the ratio of failed transmissions is the number of collisions divided by the total number of transmission attempts.
Both metrics are significantly higher for ALOHA-Q agents operating in non-saturation than under saturation, affecting achievable throughput values and explaining the non-ideal ALOHA-Q behavior presented in Figure~\ref{fig:steadystate-mid}.

To summarize, experiments with non-saturated traffic profiles demonstrate that KISS performs significantly better than the baselines.

\subsection{Dynamic Case}

\begin{figure}[!t]
    \centering

    \begin{subfigure}[b]{0.48\linewidth}
        \centering
        \includegraphics[width=\linewidth]{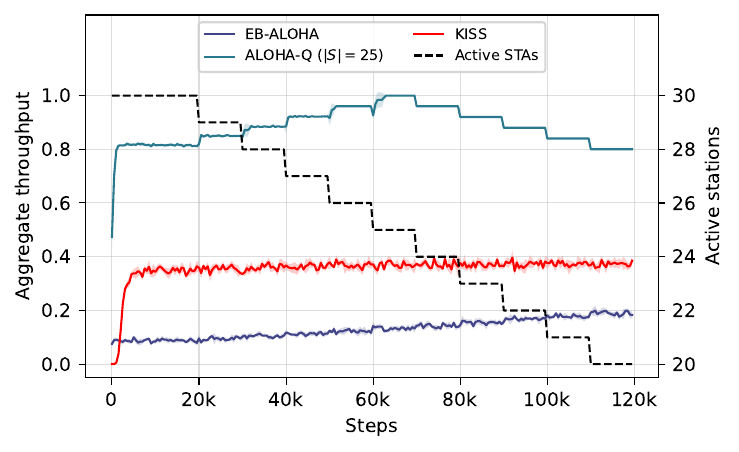}
        \caption{Throughput vs time ($N: 30 \rightarrow 20$).}
        \label{fig:dynamic-throughput-30-20}
    \end{subfigure}
    \hfill
    \begin{subfigure}[b]{0.48\linewidth}
        \centering
        \includegraphics[width=\linewidth]{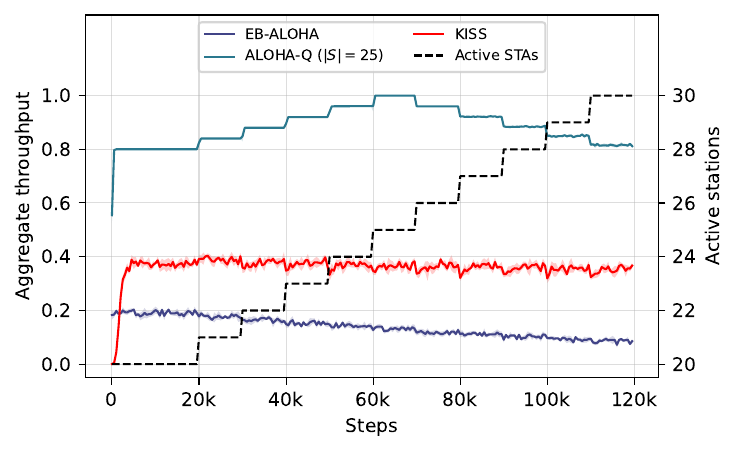}
        \caption{Throughput vs time ($N: 20 \rightarrow 30$).}
        \label{fig:dynamic-throughput-20-30}
    \end{subfigure}

    \caption{Performance for the dynamic case. }
    \label{fig:dynamic-throughput-subfigs}
\end{figure}

Finally, we examine a dynamic scenario in which stations join or leave the network over time, allowing us to assess the flexibility of each protocol. Figure~\ref{fig:dynamic-throughput-subfigs} shows the aggregate throughput for two such scenarios, one with an increasing number of stations and one with a decreasing number. In these experiments, we exclude ideal ALOHA‑Q and DLMA. Estimating the number of active stations is difficult in real deployments, so instead of ideal ALOHA‑Q, we use a fixed variant with a time window of 25, corresponding to the average network size in our dynamic tests. DLMA is also omitted due to its long exploration time and poor performance observed in earlier evaluations.

The aggregate throughput of ALOHA-Q is heavily dependent on the number of active stations. When the number of stations decreases (increases), the throughput increases (drops).
As in the steady-state case, ALOHA-Q performs best when the number of stations is close to the number of available time slots. As a result, ALOHA-Q requires continuous monitoring of the number of users and dynamic adjustment of the number of time slots to operate effectively in a dynamic environment.

The aggregate KISS throughput remains constant at $\approx0.4$, regardless of the number of stations. The network experiences slight temporary throughput drops when a single station connects or disconnects from the network. This occurs because the stations must adapt to the new environment (network state) and spend some time on exploration. Nevertheless, KISS is the only distributed channel access scheme that adapts successfully to a dynamic environment.

\section{Ablation Studies}
\label{sec:ablation}

We evaluate KISS’s sensitivity to its state representation, reward components, and action‑space constraints. Each experiment modifies only one element of the default configuration to isolate its impact.

\subsection{Observation History Length}

\begin{figure}[!t]
    \centering
    \begin{minipage}[t]{0.48\linewidth}
        \includegraphics[width=\linewidth]{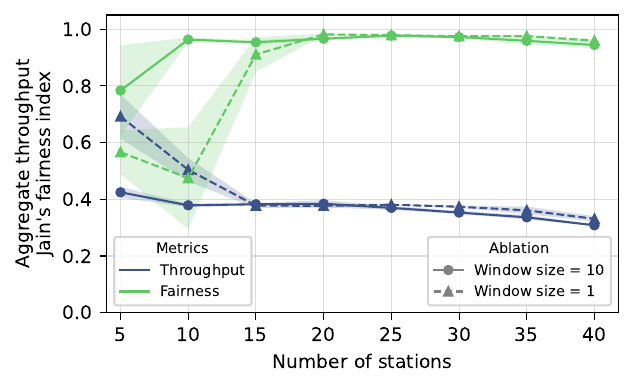}
        \caption{Ablation of the observation history length $L$. For small networks ($N \le 10$), shorter history ($L=1$) leads to higher throughput at the cost of significantly lower fairness. For larger networks ($N \ge 15$), both configurations converge to similar performance.}
        \label{fig:ablation-queue}
    \end{minipage}
    \hfill
    \begin{minipage}[t]{0.48\linewidth}
        \includegraphics[width=\linewidth]{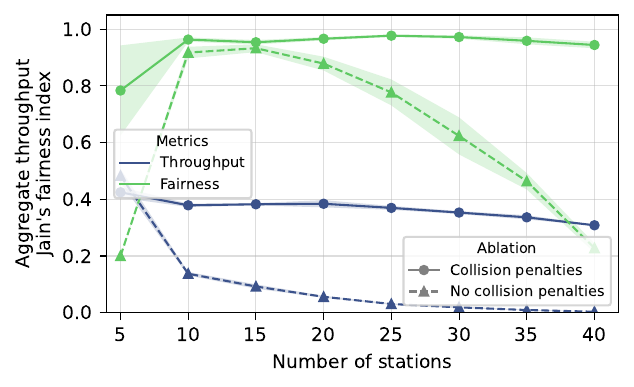}
        \caption{Ablation of the collision penalty $P_{\mathrm{coll}}$. Without it, throughput collapses to near zero for $N \ge 25$ as agents fail to coordinate.}
        \label{fig:ablation-collision}
    \end{minipage}
\end{figure}

The state representation includes a history of the last $L$ instantaneous observations. Figure~\ref{fig:ablation-queue} compares the default configuration ($L=10$) against a minimal history length ($L=1$). For $N=5$ and $N=10$, the $L=1$ configuration achieves a higher aggregate throughput but at the cost of significantly lower fairness. This indicates that without a temporal context, agents converge to more aggressive or asymmetric policies that favor channel capture by a subset of nodes. However, as the network size increases ($N \ge 15$), the performance of both configurations converges. For larger $N$, the aggregate throughput for $L=1$ drops to the same level as $L=10$, while fairness improves to near-optimal levels.

\subsection{Reward Components}

We selectively disable the following three components of the reward function.

\paragraph{Collision penalty.} Setting $P_{\mathrm{coll}} = 0$ (Figure~\ref{fig:ablation-collision}) has the most severe effect. Throughput  collapses to near zero for $N \ge 25$. Fairness remains below 0.2 across all scenarios, indicating that agents fail to learn any form of collision avoidance without a direct negative signal. This confirms that the collision penalty is the primary driver for decentralized consensus in the shared medium.

\begin{figure}[!t]
    \centering
    \begin{minipage}[t]{0.48\linewidth}
        \includegraphics[width=\linewidth]{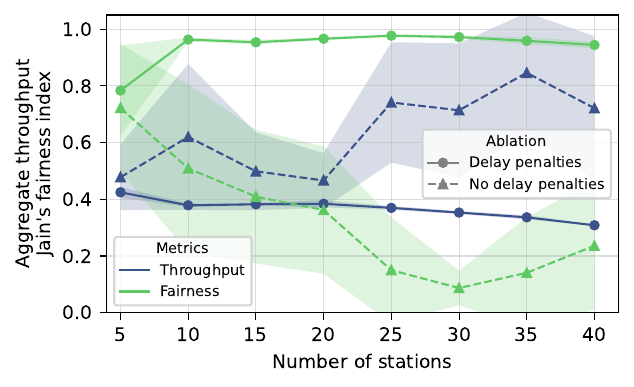}
        \caption{Ablation of the delay penalties ($S_{\mathrm{idle}}$, $D$). Removing them allows for channel capture, peaking throughput at $\approx 0.85$ but destroying fairness.}
        \label{fig:ablation-delay}
    \end{minipage}
    \hfill
    \begin{minipage}[t]{0.48\linewidth}
        \includegraphics[width=\linewidth]{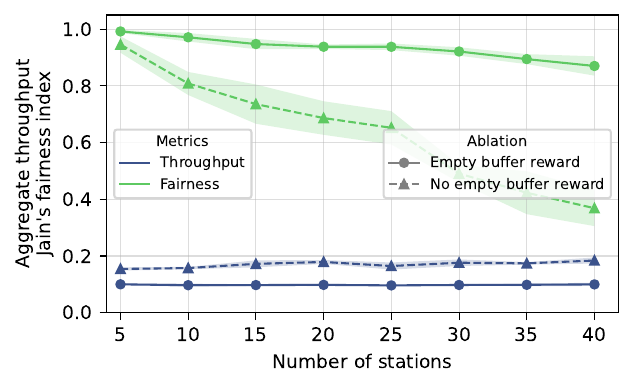}
        \caption{Ablation of the empty buffer reward. Removing it increases throughput but degrades the coordination required for high fairness in dense networks.}
        \label{fig:ablation-empty-buffer}
    \end{minipage}
\end{figure}

\paragraph{Delay penalty.} Removing delay penalties ($S_{\mathrm{idle}}$, $D$) induces a strong throughput--fairness tradeoff (Figure~\ref{fig:ablation-delay}). Aggregate throughput climbs to $\approx 0.85$ for $N=35$ (compared to the $\approx 0.35$ baseline), but Jain's index vanishes, falling below $0.1$ for $N \ge 30$. Without a penalty for idling or buffer growth, some agents adopt an aggressive channel-capture strategy that maximizes total output by starving other nodes.

\paragraph{Empty buffer reward.} Disabling rewards for maintaining an empty buffer ($R_{\mathrm{idle}}$, $P_{\mathrm{empty}}$) in a low traffic scenario increases throughput from $\approx 0.10$ to $\approx 0.18$, but fairness drops from $\approx 0.9$ to $\approx 0.38$ at $N=40$ (Figure~\ref{fig:ablation-empty-buffer}). This suggests the empty buffer reward prevents agents from monopolizing resources when their own demand is met, ensuring fair resource distribution in dense networks.

\subsection{Action Space Constraints}

We also evaluate the impact of constraining the action space of the KISS agents with hand-crafted heuristics such as listen-before-talk (LBT) and a rule that prevents transmission attempts when the buffer is empty.

\paragraph{Listen before talk.}

\begin{figure}[!t]
    \centering
    \begin{minipage}[t]{0.48\linewidth}
        \includegraphics[width=\linewidth]{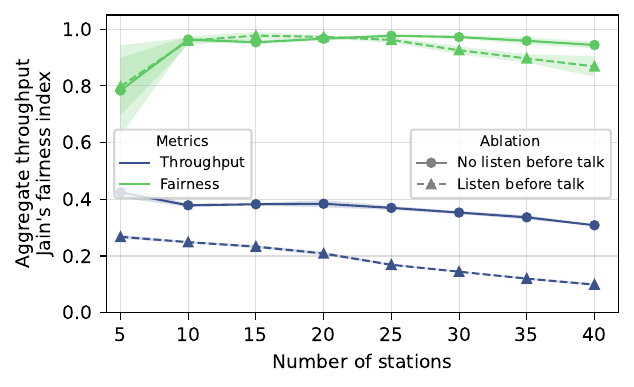}
        \caption{Ablation of listen-before-talk. LBT significantly reduces aggregate throughput and degrades fairness in larger networks compared to unconstrained KISS.}
        \label{fig:ablation-lbt}
    \end{minipage}
    \hfill
    \begin{minipage}[t]{0.48\linewidth}
        \includegraphics[width=\linewidth]{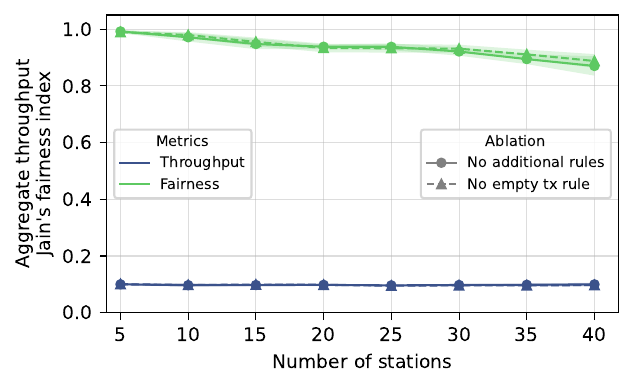}
        \caption{Ablation of empty buffer transmission rule. The performance of unconstrained KISS is almost equal to the one with a hard-coded 'no empty tx' rule, suggesting that agents autonomously learn this behavior (low traffic profile used).}
        \label{fig:ablation-no-empty-tx}
    \end{minipage}
\end{figure}

With an LBT heuristic agents must sense an idle channel before transmitting. We observe that introducing the LBT constraint significantly reduces the aggregate throughput (Figure~\ref{fig:ablation-lbt}). This reduction is caused by the fact that stations must sense the channel and wait for an idle slot, effectively introducing a mandatory overhead of at least one slot for every successful transmission. While LBT maintains relatively high fairness for small networks, its fairness starts to degrade for $N > 20$, eventually falling below the performance of unconstrained KISS agents. These results confirm that constraining the action space with a hand-crafted heuristic such as LBT is counterproductive.

\paragraph{No transmissions with empty buffer.}

We evaluate the impact of a rule that explicitly prevents transmission attempts when the buffer is empty. This constraint is theoretically beneficial as it prevents wasteful use of the radio resources. However, the performance of the constrained agent is almost identical to the unconstrained one in a low traffic scenario (Figure~\ref{fig:ablation-no-empty-tx}). This indicates that the KISS agent autonomously learns to avoid transmitting with an empty buffer, driven by the reward function, making such a hard-coded constraint redundant.

\subsection{Summary}

Our ablation studies reveal that the reward design is the most critical part for KISS operation, as removal of its components leads to severe performance degradation. Specifically, the collision penalty is the primary driver for decentralized consensus; without it, throughput collapses as coordination fails. Furthermore, delay and empty-buffer rewards are essential for maintaining Jain's fairness index by preventing aggressive channel-capture behaviors and starvation. We also find that a temporal history of observations ($L=10$) is beneficial in ensuring fairness in smaller networks, although its impact diminishes as the network density increases. Finally, we demonstrate that imposing rigid heuristics like LBT or a ``no empty tx'' rule is counterproductive or redundant. These rules either limit the agents' ability to discover more efficient stochastic transmission strategies or are autonomously learned. Our findings underscore that KISS succeeds by combining simple slot-level observations with a carefully balanced multi-objective reward function.

\section{Conclusions and Future Work}

Although many research articles address the problem of distributed and decentralized wireless network operation, we demonstrate that ML agents can autonomously learn channel‑access strategies without inheriting the shortcomings identified in the literature. In this sense, these agents should be viewed primarily as a tool for \textit{discovering} decentralized access strategies rather than as the final runtime protocol. One might argue that we employ a complex ML system to learn behavior that a simple, hand‑crafted heuristic could approximate. However, the key distinction is that our solution learns effectively under stricter and more realistic conditions than those assumed in existing ML‑based approaches. KISS is fully online (requiring no pre‑training), fully distributed (with each agent learning and acting independently), stochastic (with no reliance on periodic schedules), and requires no coordination or explicit communication to achieve efficient and fair channel access.

The proposed system model allows agents to learn a simple yet efficient slotted-ALOHA method with dynamic transmission probabilities and quickly adapt to dynamic environmental changes (e.g., node mobility). Interestingly, the agents converge to a simple slotted-ALOHA-like strategy with dynamically adjusted transmission probabilities, suggesting that decentralized learning can autonomously rediscover efficient random access mechanisms.
Furthermore, comprehensive ablation studies led to distillation of the most impactful state representations, reward components, and action space constraints.

In our studies, we have so far assumed a simplified traffic generation model. Future work is required for more realistic traffic patterns, such as bursts.
Additionaly, more realistic network conditions should be considered, including propagation delays, capture effect, hidden nodes, variable packet lengths, energy consumption, and potential sources of external interference.
Finally, ML interpretability is an important current trend that improves its trustworthiness. Therefore, we plan to use symbolic regression to extract the protocol learned by the neural network in symbolic form.
Another direction for future work is to include on‑policy baselines, such as multi-agent proximal policy optimization (MAPPO).

\section*{Acknowledgments}
This research was funded by the National Science Centre, Poland (2023/05/Y/ST7/00004). For the purpose of Open Access, the authors have applied a CC-BY public copyright licence to any Author Accepted Manuscript (AAM) version arising from this submission. We gratefully acknowledge the support of the National Research Institute, grant number POIR.04.02.00-00-D008/20-01 on ``National Laboratory for Advanced 5G Research'' (acronym PL-5G) for providing computing facilities.

\bibliographystyle{unsrtnat}
\bibliography{references.bib}

\end{document}